\documentclass[12pt]{article}

\usepackage{scicite,epsfig,amssymb,amsmath,lscape}

\usepackage{times}
\usepackage{caption}
\usepackage{color}

\usepackage[pdftex,breaklinks]{hyperref}

\newcommand{\begit}{\begin{itemize}}
\newcommand{\enit}{\end{itemize}}
\newcommand{\begen}{\begin{enumerate}}
\newcommand{\enen}{\end{enumerate}}

\newcommand{\beq}{\begin{equation}}
\newcommand{\eeq}{\end{equation}}
\newcommand{\beqa}{\begin{eqnarray}} 
\newcommand{\eeqa}{\end{eqnarray}}

\newdimen\sa  \newdimen\sb
\def\parcs{\sa=.07em \sb=.03em
     \ifmmode $\rlap{.}$^{\scriptscriptstyle\prime\kern -\sb\prime}$\kern -\sa$
     \else \rlap{.}$^{\scriptscriptstyle\prime\kern -\sb\prime}$\kern -\sa\fi}

\newenvironment{sciabstract}{%
\begin{quote} \bf}
{\end{quote}}

\newcounter{lastnote}

\title{Response to Comment on ``A Non-Interacting Low-Mass Black Hole$-$Giant Star Binary System"}

\author
{Todd A.\ Thompson,$^{1,2\ast}$ 
Christopher S.\ Kochanek,$^{1,2}$
Krzysztof Z.\ Stanek,$^{1,2}$\\   
Carles\ Badenes,$^{3,4}$ 
Tharindu Jayasinghe,$^{1,2}$ 
Jamie Tayar,$^{5}$
Jennifer A.\ Johnson,$^{1,2}$\\
Thomas W.-S.\ Holoien,$^{6}$ 
Katie Auchettl,$^{2,7,8}$
Kevin Covey$^{9}$}

\date{}

\begin{document} 

% Double-space the manuscript.

\baselineskip24pt

% Make the title.

\maketitle 

\noindent
\normalsize{$^{1}$ Department of Astronomy, The Ohio State University, 140 W.\ 18th Ave., Columbus, OH 43210, USA}\\
\normalsize{$^{2}$ Center for Cosmology and AstroParticle Physics, The Ohio State University, 191 W.\ Woodruff Ave., Columbus, OH 43210, USA } \\
\normalsize{$^{3}$Department of Physics and Astronomy and Pittsburgh Particle Physics, Astrophysics and Cosmology Center (PITT PACC), University of Pittsburgh, 3941 O'Hara Street, Pittsburgh, PA 15260, USA}\\
\normalsize{$^{4}$  Institut de Ci\`encies del Cosmos, Universitat de Barcelona, Mart\'i Franqu\'es 1, E08028 Barcelona, Spain} \\
\normalsize{$^{5}$ Institute for Astronomy, University of Hawaii, 2680 Woodlawn Drive, Honolulu, Hawaii 96822, USA \\
\normalsize{$^{6}$  The Observatories of the Carnegie Institution for Science, 813 Santa Barbara St., Pasadena, CA 91101, USA}\\
\normalsize{$^{7}$ Department of Physics, The Ohio State University, 191 W. Woodruff Avenue, Columbus, OH 43210, USA}  \\
\normalsize{$^{8}$ DARK, Niels Bohr Institute, University of Copenhagen, Lyngbyvej 2, 2100 Copenhagen, Denmark}\\
\normalsize{$^{9}$ Department of Physics and Astronomy, Western Washington University, Bellingham, WA, 98225, USA} \\
\normalsize{$^\ast$To whom correspondence should be addressed; E-mail: thompson.1847@osu.edu}

\baselineskip24pt
\begin{sciabstract}
van den Heuvel \& Tauris argue that if the red giant star in the system 2MASS J05215658+4359220 has a mass of $1$\,solar mass ($M_\odot$), then its unseen companion could be a binary composed of two 0.9\,M$_\odot$ stars, making a triple system. We contend that the existing data are most consistent with a giant of mass $3.2^{+1.0}_{-1.0}$\,M$_\odot$, implying a black hole companion of $3.3^{+2.8}_{-0.7}$\,M$_\odot$.
\end{sciabstract}

van den Heuvel \& Tauris \cite{vandenheuvel} posit that the red giant star in the system 2MASS J05215658 +4359220 \cite{Thompson2019} could have a mass of $M_{\rm giant}\simeq1$\,M$_\odot$, and that the unobserved companion could be a normal stellar binary system composed of two 0.9\,M$_\odot$ stars. This hypothesis is inconsistent with the measured luminosity $L$ and effective temperature $T_{\rm eff}$. The latter was established by three independent and consistent measurements: (a) the optical spectra, (b) the near-infrared spectra, and (c) the fit to the giant's spectral energy distribution (SED). The luminosity was determined from two independent methods: (a) the observed SED combined with the measured distance, and (b) the stellar radius ($R$) as inferred from the giant's projected rotational velocity ($v\sin i$) combined with $T_{\rm eff}$. Both yield consistent $L$ for $\sin i\simeq1$. Given these data and their uncertainties, and acknowledging the inherent systematic uncertainties in comparing with evolutionary models, we disfavored $M_{\rm giant}\simeq1$\,M$_\odot$, and obtained a best-fitting value of $M_{\rm giant}\simeq3.2^{+1.0}_{-1.0}$\,M$_\odot$ (2-$\sigma$ uncertainties) \cite{Thompson2019}. 

van den Heuvel \& Tauris  assert that ``Spectroscopic determination of a red giant's mass from model atmospheres can be uncertain by a factor of 3."  However, we do not determine the mass from the logarithm of the stellar gravitation acceleration ($\log g$) alone using $10^{\log g}=GM_{\rm giant}/R^2$, but instead from fitting $L$, $T_{\rm eff}$, and $\log g$ to evolutionary models. Even ignoring the constraint on $\log g$, the combination of $L$ and $T_{\rm eff}$ are inconsistent with $M_{\rm giant}\simeq1$\,M$_\odot$. The mass obtained from $M_{\rm giant}= R^2 10^{\log g}/G$ is consistent with our best-fitting $M_{\rm giant}$, but it is not the origin of our final reported mass.

van den Heuvel \& Tauris argue that because X-ray emission is seen in symbiotic X-ray binary systems, it should be seen in 2MASS J05215658+4359220 if the unseen companion is a black hole. We do not find this argument convincing. First, the expected X-ray emission depends on the mass loss rate of the giant, which is uncertain. In particular, some studies have found lower values of the mass loss rate normalization than used by van den Heuvel \& Tauris  \cite{Miglio2012,Choi2016}. Second, the expected accretion rate is strongly dependent on the assumed giant wind velocity, which is not well-constrained for this system. Third, the expected X-ray emission depends on the radiative efficiency of the accretion and the nature of the accretor. We estimated the accretion rate and found that the system may be in the radiatively inefficient regime \cite{Thompson2019}, implying low X-ray luminosity. In addition, the dichotomy between the X-ray luminosities of neutron star- and black hole-hosting Galactic X-ray binaries in quiescence may result from the presence of an event horizon for black holes, whereas neutron stars have surfaces \cite{Menou1999,Garcia2001,McClintock2004}. The lack of X-ray emission observed in 2MASS J05215658+4359220 may then be used to argue for a black hole companion instead of a neutron star. In fact, the X-ray emission in the symbiotic X-ray binary systems discussed by van den Heuvel \& Tauris are often explained using a settling accretion flow model relevant to neutron stars and not black holes (\cite{Yungelson2019}; their Section 2). 

van den Heuvel \& Tauris state that ``the $[{\rm C/N}]$ ratio of the giant would be unusually high for giants of this mass" and that ``the high $[{\rm C/N}]$  abundance ratio is normal for a 1\,M$_\odot$ red giant." While the measured abundance ratio is somewhat unusual for a $\simeq3$\,M$_\odot$ giant in our comparison sample, whether it is unusual enough to outweigh the well-measured values of $L$ and $T_{\rm eff}$ is debatable. Above 3\,M$_\odot$, we found that 1 out of 18 stars (about 6\%) have high $[{\rm C/N}]$ \cite{Thompson2019}. While the observation of $[{\rm C/N}]\simeq0.0$ for 2MASS J05215658+4359220 might be used to argue that $M_{\rm giant}\simeq1$\,M$_\odot$ in the absence of any other information, the additional information provided by $L$ and $T_{\rm eff}$ indicates a significantly more massive star when compared to evolutionary models with a variety of metallicities \cite{Thompson2019}. Given the spotted, rapidly rotating nature of the giant we cannot exclude systematic uncertainties in the determination of both the $[{\rm C/N}]$ abundance ratio and the metallicity. The latter affects the fitting of the giant to evolutionary models \cite{Thompson2019}. Systematic uncertainties of $\pm0.1-0.3$\,dex for C, $\pm0.2$\,dex for N, and $\pm0.1$\,dex for Fe are found when APOGEE abundances are compared with other determinations from the literature \cite{Jonsson2018}. 

The proposal by van den Heuvel \& Tauris is also inconsistent with the limits on ellipsoidal variability derived from the ASAS-SN light curve (\cite{Thompson2019}; Section 1.5.5, Supplementary Material). This is because the system would have lower total mass, but the same orbital period, so the semi-major axis would decrease relative to the radius of the giant, and the mass ratio would shift to be more dominated by the putative binary companion. These limits on ellipsoidal variability could be avoided if the distance to the system is smaller, so that the giant star has lower luminosity and smaller radius, while keeping $\sin i\simeq1$, but the required change in distance is inconsistent with the parallax uncertainties, and the observed $v \sin i$ would then no longer be consistent with the stellar radius.

We conclude that the hypothesis of van den Heuvel \& Tauris is inconsistent with the measurements of $L$, $T_{\rm eff}$, $\log g$, and $v \sin i$ and that the low X-ray luminosity may be accommodated by a black hole companion.

\bibliographystyle{Science}
\bibliography{response.bib}

\begin{thebibliography}{10}

\bibitem{vandenheuvel}
E.~P.~J. {van den Heuvel}, T.~M. {Tauris}, {\it Science, {\rm this issue}\/}
  (2020).

\bibitem{Thompson2019}
T.~A. {Thompson}, {\it et~al.\/}, {\it Science\/} {\bf 366}, 637 (2019).

\bibitem{Miglio2012}
A.~{Miglio}, {\it et~al.\/}, {\it \mnras\/} {\bf 419}, 2077 (2012).

\bibitem{Choi2016}
J.~{Choi}, {\it et~al.\/}, {\it \apj\/} {\bf 823}, 102 (2016).

\bibitem{Menou1999}
K.~{Menou}, {\it et~al.\/}, {\it \apj\/} {\bf 520}, 276 (1999).

\bibitem{Garcia2001}
M.~R. {Garcia}, {\it et~al.\/}, {\it \apjl\/} {\bf 553}, L47 (2001).

\bibitem{McClintock2004}
J.~E. {McClintock}, R.~{Narayan}, G.~B. {Rybicki}, {\it \apj\/} {\bf 615}, 402
  (2004).

\bibitem{Yungelson2019}
L.~R. {Yungelson}, A.~G. {Kuranov}, K.~A. {Postnov}, {\it \mnras\/} {\bf 485},
  851 (2019).

\bibitem{Pinsonneault18}
M.~H. {Pinsonneault}, {\it et~al.\/}, {\it \apjs\/} {\bf 239}, 32 (2018).

\bibitem{Jonsson2018}
H.~{J{\"o}nsson}, {\it et~al.\/}, {\it \aj\/} {\bf 156}, 126 (2018).

\end{thebibliography}

\section*{Acknowledgements}
{\bf Acknowledgments:} T.A.T.~thanks K.~A.~Byram for discussions and support.
{\bf Funding:} T.A.T. acknowledges support from NASA grant \#80NSSC20K0531.  C.B. acknowledges support from PHY 14-30152: Physics Frontier Center/JINA Center for the Evolution of the Elements (JINA-CEE), awarded by the National Science Foundation, from Scialog Scholar grant 24216 from the Research Corporation, and from NSF grant AST 19-09022.  C.S.K. and K.Z.S. are supported by NSF grants AST-1515927, AST-181440, and  AST-1907570. J.T. acknowledges support from NASA through the NASA Hubble Fellowship grant \#51424 awarded by the Space Telescope Science Institute, which is operated by the Association of Universities for Research in Astronomy, Inc., for NASA, under contract NAS5-26555.
{\bf Author contributions:} T.A.T. led the interpretation of the system and the writing of the response. C.S.K.,  K.Z.S., C.B., and T.J. consulted on aspects of the system, including evolutionary model fitting and ellipsoidal variations, and edited the text.  J.T. and  J.A.J. provided input on the text and on the analysis of the APOGEE spectra and the abundances. T.W.S.H., K.A., and K.C. participated in readying the text for publication. 
{\bf Competing interests:} The authors declare that there are no competing interests.  
{\bf Data and materials availability:} There are no new data.

\end{document}